\documentclass[aps,prl,reprint,amsmath,amssymb,groupedaddress]{revtex4-2}
\usepackage{graphicx}
\usepackage[colorlinks,
linkcolor=blue,
anchorcolor=blue,
citecolor=blue,urlcolor=blue]{hyperref}
\begin{document}
	
	
	\title{A general multi-wave quasi-resonance theory for lattice energy diffusion}
	
	
	\author{Wei Lin$^1$}
	\author{Weicheng Fu$^{2,3}$}
	\author{Zhen Wang$^4$}
	\author{Yong Zhang$^{1,3}$}
	\author{Hong Zhao$^{1,3}$}
	\email{E-mail: zhaoh@xmu.edu.cn}
	
	\affiliation{$^1$Department of Physics, Xiamen University, Xiamen 361005, Fujian, China\\ 
		$^2$Department of Physics, Tianshui Normal University, Tianshui 741001, Gansu, China\\
		$^3$Lanzhou Center for Theoretical Physics, Key Laboratory of Theoretical Physics of Gansu Province, Lanzhou University, Lanzhou, Gansu 730000, China\\
		$^4$CAS Key Laboratory of Theoretical Physics and Institute of Theoretical Physics,Chinese Academy of Sciences, Beijing 100190, China}
	
	
	
	\date{\today}
	
	\begin{abstract}
		In this letter, a multi-wave quasi-resonance framework is established to analyze energy diffusion in classical lattices, uncovering that it is fundamentally determined by the characteristics of eigenmodes. Namely, based on the presence and the absence of extended modes, lattices fall into two universality classes with qualitatively different thermalization behavior. In particular, we find that while the one with extended modes can be thermalized under arbitrarily weak perturbations in the thermodynamic limit, the other class can be thermalized only when perturbations exceed a certain threshold, revealing for the first time the possibility that a lattice cannot be thermalized, violating the hypothesis of statistical mechanics. Our study addresses conclusively the renowned Fermi-Pasta-Ulam-Tsingou problem for large systems under weak perturbations, underscoring the pivotal roles of both extended and localized modes in facilitating energy diffusion and thermalization processes.
		
	\end{abstract}
	
	
	\maketitle 
	
	Thermalization or energy equipartition assumption is a cornerstone for applying statistical physics and thermodynamics theories and also the premise for the zeroth law of thermodynamics. The validation of the equipartition theorem for gas models is shown in the framework of ergodic theory and supported by subsequent experimental findings. In contrast, whether a lattice model universally adheres to this theorem and if so, what the specific conditions are required, have persisted as a research focus since the seminal numerical simulation Fermi's team performed in 1953~\cite{Fermi:1955}. This pioneer effort put forward the celebrated Fermi-Pasta-Ulam-Tsingou (FPUT) problem, fostering the development of diverse physics disciplines such as nonlinear dynamics and chaos, integrability and solitary waves, and computational physics. In fact, significant strides have been made for lattice models as well~\cite{Fermi:1955, Zabusky:1965, Fucito:1982, Luca:1995, Poggi:1995, Parisi:1997, Casetti:1997, Luca:1999, Berchialla:2004, Gershgorin:2007, Gallavotti:2008, Benettin:2009, Cretegny:1998, Lepri:1998, Benettin:2011, Benettin:2013, Zhang:2016, Sun:2020, Ganapa:2020}, but as most of them are reached by pure numerical simulations, unambiguous conclusions are still lack.
	
	The recent integration ~\cite{Onorato:2015, Lvov:2018, Pistone:2018, Fu:2019R, Fu:2019, Pistone:2019, Onorato:2023, Wang:2020, Wang:2020a} of wave turbulence theory~\cite{Zakharov:2004,Nazarenko:2011} into the FPUT problem has created a theoretical analysis framework with a complete logical chain. This methodology treats the normal modes of the lattice Hamiltonian as waves, establishing connections between energy diffusion and multi-wave resonances under specific frequency and wave vector combinations. It ensures energy redistribution across all modes by demonstrating the presence of a connected network of multi-wave resonances, thereby facilitating energy transfer throughout the entire system. This systematic approach is referred to as the exact resonance approach. Notably, these pioneer researchers have demonstrated that for the FPUT lattice models, originally explored by Fermi, Pasta, Ulam, and Tsingou with only a few dozen of particles, a 6-wave resonance network emerges, leading to energy equipartition after a sufficiently long evolution time. This methodology has sparked investigations into lattice thermalization in the thermodynamic limit~\cite{Pistone:2018, Fu:2019, Fu:2019R, Pistone:2019, Wang:2020, Wang:2020a, Onorato:2023}.
	
	However, the exact resonance approach needs be developed further. First,  numerical studies have shown that for large one-dimensional homogeneous lattices, the thermalization time scales according to 4-wave resonances,  instead~\cite{Pistone:2018, Fu:2019, Fu:2019R, Pistone:2019, Wang:2020, Wang:2020a, Onorato:2023}, while the existing exact resonance approach cannot show the formation of connected networks dominated by 4-wave exact resonances. Second, exact resonances are not fit for disordered lattices, of which the numerical studies suggest that the may follow the thermalization scaling law of 3-wave resonances~\cite{Wang:2020}. Therefore, a generalized approach that deals with both homogeneous and disordered lattices, as well as both small and large lattices on the other hand, is highly expected.
	
	Addressing the limitations of the exact resonance approach, the concept of quasi-resonance has been proposed, which is defined as transient multi-wave resonances induced by frequency broadening or shifts resulting from nonlinear interactions. This phenomenon causes mode frequencies to fluctuate around their predicted dispersion relation values, temporarily fulfilling resonance conditions. The conventional wave turbulence theory has highlighted the inadequacies of exact resonances in explaining certain phenomena and has emphasized and considered the essential role of quasi-resonance~\cite{Connaughton:2001, Kartashova:2007, Lvov:2010}. It is also qualitatively used to explain why in larger lattices, sets of four or three resonance waves can result in connected networks~\cite{Pistone:2018, Fu:2019, Fu:2019R, Pistone:2019, Wang:2020, Wang:2020a, Onorato:2023}.
	
	The purpose of this Letter is to develop an analysis framework that can quantitatively deal with the quasi-resonant effect. Our framework encompasses a pair of criteria to quantify the quasi-resonance condition, alongside a metric for evaluating the connectivity strength of the resonance network. Consequently, our approach is applicable to both homogeneous and disordered lattices, and can identify when a multi-wave quasi-resonance network predominantly controls energy diffusion and when it loses the control. This leads to the finding that there are two distinct universality classes based on the presence and the absence of extended mode. The cases that either exhibits fully extended modes~\cite{Fermi:1955, Zabusky:1965, Fucito:1982, Luca:1995, Poggi:1995, Parisi:1997, Casetti:1997, Cretegny:1998, Lepri:1998, Luca:1999, Berchialla:2004, Gershgorin:2007, Benettin:2009, Gallavotti:2008, Benettin:2011, Benettin:2013, Onorato:2015, Zhang:2016, Lvov:2018, Fu:2019, Fu:2019R, Pistone:2018, Pistone:2019, Ganapa:2020, Onorato:2023} or a combination of extended and localized modes~\cite{Wang:2020, Wang:2020a, Sun:2020} belong to the first class, where the thermalization time scales as $T_{\text{eq}} \propto g^{-2}$ for sufficiently large lattices, in spite of the deviation in comparatively smaller lattices, where $g$ represents the nonlinearity strength of the system. The thermalization of the second class characterized exclusively by localized modes is addressed for the first time here. We find that in this case the thermalization time follows the scaling of $T_{\text{eq}} \propto g^{-n}$, where $n$ increases as $g$ decreases. As there are no finite-size effects in thermalization, this class cannot be thermalized under arbitrarily weak perturbations in principle.
	
	We utilize the $\phi_4$ lattice model as our illustrative example. This model finds wide applications in various fields~\cite{Alfimov:2014,Kevrekidis:2016,Xiong:2018,Manton:2021}. Importantly, it is capable of displaying fully extended, completely localized, and mixed modes by changing its control parameters, and thus can serve as a representative model of general lattices. We stress that the conclusions drawn in this paper are not reliant on a particular lattice model, as our approach solely focuses on analyzing eigenmodes. The thermalization behavior of homogeneous $\phi_4$ lattice has been investigated previously~\cite{Pistone:2018,Pistone:2019}, but the focus has been on the phase of fully extended modes, as we will see later. The Hamiltonian of this model is expressed as
	\begin{eqnarray}\label{eq:H}
		H = \sum_{j} \left[ \frac{p_j^2}{2m_j}  + \frac{1}{2}(q_{j+1}-q_j)^2 +
		\frac{1}{2}b q_j^2 + \frac{1}{4}\beta q_j^4 \right],
	\end{eqnarray}
	where $m_j$, $p_j$, and $q_j$ represent the mass, momentum, and displacement of the $j$-th particle, respectively. The mass $m_j$ is uniformly distributed within the range $1-\delta m < m_j < 1+\delta m$, with $\delta m$ indicating the degree of disorder. The last term in the Hamiltonian introduces nonlinear interactions, characterized by $\beta$ as its amplitude. In the special case of $\beta=0$, the model becomes integrable. 
	
	We rescale $q_j$ by energy density $\epsilon$ : $q_j={\tilde q_j\epsilon^{1/2}}$ and obtain ${\tilde H}=H/\epsilon=H_0(\tilde q_j,\tilde p_j)+\sum_j \frac{1}{4}g {\tilde q_j}^{4}$, where $ g=\beta\epsilon$ represents the nonlinearity strength. In Fourier space, the Hamiltonian is transformed into ${\tilde H} = \frac{1}{2} \sum_{k} \left(P^2_k + \omega_k^2 Q_k^2 \right) + \frac{g}{4} \sum_{k_1,..,k_4} Q_{k_1}Q_{k_2}Q_{k_3}Q_{k_4} \sum_j \frac{u^{k_1}_ju^{k_2}_ju^{k_3}_ju^{k_4}_j}{m^2_j}$, with $P_k=\sum_j \tilde p_ju^k_j/\sqrt{m_j} $ and $Q_k=\sum_j \sqrt{m_j}\tilde q_j u^k_j$ representing the superpositions of the $k$-th normal mode, and $\omega_k$ denoting its frequency. Here, the normal modes are indexed by $k$, arranged in ascending order of frequency. The term $u^k_j$ refers to the renormalized amplitude of the $k$-th mode at position $j$. According to wave turbulence theory, the quartic polynomial terms imply 4-wave interactions among the normal modes. 
	
\begin{figure}[tb] 
		\includegraphics{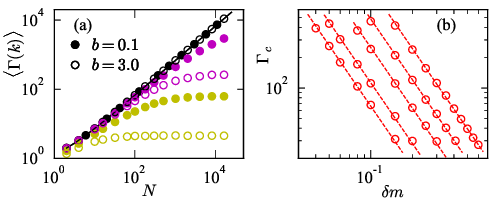}
		\caption{\label{fig:can} Scaling relationships. (a) Displays the mean participation number, $\langle\Gamma(k)
			\rangle$, as a function of $N$ for two different values of $b$: $b=3$ (represented by circles) and $b=0.1$ (indicated by dots). This is shown for three $\delta m$ values: 0.0, 0.05, and 0.5 (arranged from top to bottom). (b) Illustrates the saturated participation number, $\Gamma_c$, plotted against $\delta m$ for various on-site potential heights (from left to right: $b=3, 2, 1, 0.5, 0.2$). Dashed lines are included for visual clarity. It reveals the scaling law $\Gamma_c \propto \delta m^{-2} b^{-0.76}$, which is particularly accurate for smaller values of $b$.}
	\end{figure}

	We employ the standard participation number
	\begin{eqnarray}\label{eq:can}
		\Gamma(k) = \frac{\left[\sum_{j} \left(u^k_j\right)^2\right]^2}{\sum_j (u^k_{j})^4 } 
	\end{eqnarray}
	to measure localization~\cite{Goldhirsch:1991,Leitner:2001}. It is known that the participation number is proportional to the Anderson localization length $\xi$~\cite{Girvin:2019}, i.e., $\xi\sim\Gamma$. Fig.~\ref{fig:can}(a) displays the average $\Gamma(k)$ plotted against the lattice size $N$ for $b=3.0$ and $0.1$, under various levels of disorder. In the absence of disorder ($\delta m = 0$), the average $\Gamma(k)$ scales linearly with $N$, a relationship that can be analytically verified, part I in supplementary material (SM)~\cite{SM}. This indicates that the modes are extended in homogeneous lattices. In contrast, when $\delta m \neq 0$, the average $\Gamma(k)$ reaches a plateau,  $\Gamma_c$, as $N$ increases. The plateau indicates that all normal modes become localized. In lattices with sizes below the plateau threshold, extended and localized modes coexist. 
	
	Moreover, we observe a scaling relationship linking $\Gamma_c$ with both the disorder strength and the on-site potential height, characterized as $\Gamma_c \propto \delta m^{-2} b^{-0.76}$ for relatively small values of $b$, as illustrated in Fig.~\ref{fig:can}(b). This relationship provides insights into how the localization length varies in response to changes in the disorder degree and on-site potential height. Note that the specific scaling of $\xi \propto \delta m^{-2}$ is a widely recognized phenomenon in disordered systems~\cite{Kramer:1993,Ishii:1973,Krimer:2010}.
	
	We present our quasi-resonance approach as follows. Initially, we derive the kinetic equations for a normal mode:
	\begin{eqnarray}\label{eq:4keq}
		\dot{D}_{k_1} = && -2 g^2 \pi \sum_{k_2,k_3,k_4} A^2_{k_1k_2k_3k_4} (C_{k_1k_2k_3k_4}\omega_{k_1k_2k_3k_4}  \nonumber  \\
		&&  + 3C_{k_1k_2k_3}^{k_4}\omega_{k_1k_2k_3}^{k_4}+ 3C^{k_3k_4}_{k_1k_2}\omega_{k_1k_2}^{k_3k_4}\nonumber  \\
		&&+  C^{k_2k_3k_4}_{k_1}\omega_{k_1}^{k_2k_3k_4})
	\end{eqnarray}
	Here, $D_{k_1}=\langle a_{k_1} a^*_{k_1} \rangle$, $a_k$ is the complex mode defined as $a_{k_1} = \frac{1}{\sqrt{2\omega_{k_1}}}\left(\omega_{k_1} Q_{k_1} + iP_{k_1}\right)$, and thus the mode $k_1$'s energy is $E_{k_1} = D_{k_1}\omega_{k_1}$. The 4-wave resonance amplitude is expressed as $A_{k_1k_2k_3k_4}= \frac{1}{4 \sqrt{\omega_{k_1} \omega_{k_2} \omega_{k_3} \omega_{k_4}}} \sum_j \frac{ u^{k_1}_ju^{k_2}_ju^{k_3}_ju^{k_4}_j}{m^2_j}$. In the brackets on the right-hand side of Eq.~\eqref{eq:4keq}, $C^{k_3k_4}_{k_1k_2}=D_{k_1} D_{k_2} D_{k_3} D_{k_4}\left(\frac{1}{D_{k_3}} + \frac{1}{D_{k_4}}-\frac{1}{D_{k_1}}-\frac{1}{D_{k_2}} \right)$ and $\omega_{k_1k_2}^{k_3k_4}= \delta ( \omega_{k_3} + \omega_{k_4} - \omega_{k_1} - \omega_{k_2})$. Other terms in the brackets can be expressed similarly. The equation simplifies to
	\begin{eqnarray}\label{eq:eta}
		\frac{dD_{k_1}}{dt} = \eta_{1} - \gamma_{1}D_{k_1},
	\end{eqnarray}
	where $\eta_{1}$ and $\gamma_{1}$ are independent of $D_{k_1}$ and both are proportional to $g^2$\cite{Wang:2020,Onorato:2023}. This equation leads to a relaxation time scaling of $1/\gamma_{1} \sim g^{-2}$ for mode $k_1$ if $\gamma_{1}$ is a non-vanishing constant (see part II, SM ~\cite{SM} for details).  
	
	For a homogeneous lattice, one can determine if $\gamma_{1}$ is non-vanishing by checking the resonance conditions for wave vectors and frequencies:
	\begin{equation}\label{eq:k}
		k_1 \pm k_2  \pm k_3 \pm k_4 \ \ \mathrm{mod} \ \ N\  = \ 0,
	\end{equation}
	\begin{equation}\label{eq:omega}
		\omega_{k_1} \pm \omega_{k_2} \pm \omega_{k_3} \pm \omega_{k_4} = 0,
	\end{equation}
	However, these criteria do not apply to disordered lattices. We suggest using the following criterion to uniformly ascertain whether $\gamma_{1}$ is non-zero in both types of lattices:
	
	\begin{equation}\label{eq:zhun}
		\lvert  \omega_{k_1} +  \omega_{k_2} -  \omega_{k_3} -  \omega_{k_4} \rvert < \Omega  \ \ \& \ \ A_{k_1k_2k_3k_4} \ne 0.
	\end{equation}
	The logic behind this criterion is as follows: Consider a set of modes from $H_0$ where $ \omega_{k_1} \pm  \omega_{k_2} \pm  \omega_{k_3} \pm  \omega_{k_4}  \ne 0$, indicating non-compliance with exact resonance conditions for frequencies. Perturbing $H_0$ with a nonlinear term $H'$ induces frequency shifts or broadening in $H_0$'s eigenmodes, as per nonlinear dynamics theory ~\cite{Gershgorin:2007, Lvov:2018}. This results in temporary, time-dependent frequencies. Using instantaneous frequencies instead of $H_0$'s, the condition $\omega'_{k_1}(t) \pm \omega'_{k_2}(t) \pm \omega'_{k_3}(t) \pm \omega'_{k_4}(t) = 0$ can be temporarily satisfied. In terms of $H_0$'s frequencies, this leads to the quasi-resonance condition $\lvert  \omega_{k_1} \pm  \omega_{k_2} \pm  \omega_{k_3} \pm  \omega_{k_4} \rvert  \le \Omega$ ~\cite{Connaughton:2001, Kartashova:2007, Lvov:2010}, where $\Omega$ is a constant related to frequency broadening determined by the nonlinearity strength.
	
	Nonlinearity can only induce small changes in frequencies, and for a given nonlinearity strength, the boundary of $\Omega$ remains fixed. This fact implies that specific combinations like $\lvert \omega_{k_1}+\omega_{k_2}-\omega_{k_3}-\omega_{k_4} \rvert \le \Omega$ are more achievable. This is because one can choose pairs of neighboring frequencies to construct $\omega_{k_1}-\omega_{k_3}$ and $\omega_{k_2}-\omega_{k_4}$. Each pair is the order of magnitude of the frequency interval, and their combination may be therefore an order of magnitude smaller than the frequency interval itself. Conversely, combinations like $ \omega_{k_1}+\omega_{k_2}+\omega_{k_3}-\omega_{k_4} $ are less likely, as their residuals are generally larger than the frequency interval. Hence, for identifying dominant resonances, our focus is primarily on combinations like $\omega_{k_1} + \omega_{k_2} - \omega_{k_3} - \omega_{k_4}$, which also represent the principal resonances in homogeneous lattices as noted in~\cite{Pistone:2019}.
	
	Meeting the frequency quasi-resonance condition doesn't automatically guarantee that $\gamma_{1} \ne 0$. We must also verify if $A_{k_1k_2k_3k_4} \ne 0$. This condition is checked by examining whether the resonance condition for wave vectors (Eq.~(\ref{eq:k})) is satisfied in the conventional resonance approach, which is only applicable for homogeneous lattices. Here, in order to also provide a judgment in the case of disordered lattices, we directly calculate its value. $A_{k_1k_2k_3k_4}$ represents overlap integrals between four normal modes. For extended modes, its value is determined by the integral over the entire lattice. When localized modes are involved, only the integral within the smallest localization length contributes to the overlap. 
	
	\begin{figure}[tb] 
		\includegraphics{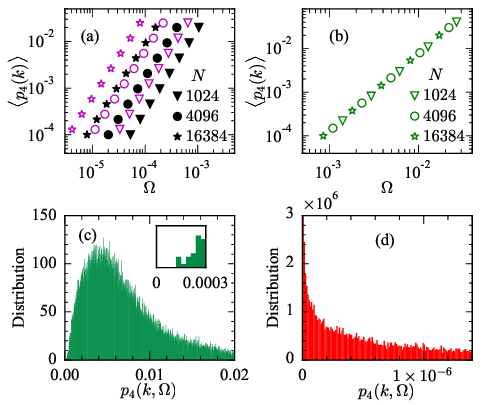}
		\caption{\label{fig:zhun} Connection strength properties represented by $p_4(k)$. (a) Illustrates $\langle p_4(k) \rangle$ versus $\Omega$ for lattices with entirely extended modes at $b=3$ and $\delta m=0$, shown with solid symbols, and those with partially extended modes at $b=0.1$ and $\delta m=0.05$, indicated by hollow symbols. (b) Depicts $\langle p_4(k) \rangle$ for lattices with completely localized modes at $b=3$ and $\delta m=0.5$. Panels (c) and (d) display the $p_4(k)$ distributions under the parameters $b=3$, $\delta m=0.5$, and $N=16384$, for $\Omega=0.01$ and $\Omega=0.0001$, respectively. The inset in (c) provides a magnified view, highlighting a distinct gap near the origin.}
	\end{figure}
	
	Another key step in our approach is the introduction of connection strength:
	\begin{equation}\label{eq:p4}
		p_4(k) = g\sum_{k_2, k_3, k_4} \left| A_{kk_2k_3k_4} \right|
	\end{equation}
	This serves as an assessment of the energy relaxation capability of the $k$-th mode to other modes. All combinations included in the summation satisfy criterion (\ref{eq:zhun}). The relationship between $g$ and $\Omega$ will be given later. The value of $p_4(k)$ is approximately proportional to the number of resonance sets connected to the $k$-th mode. A $p_4(k)$ value of 0 signifies the disconnection of this mode from the resonance network, whereas a large $p_4(k)$ indicates its connection to a substantial number of other modes, thereby facilitating the distribution of its energy throughout the network. The connection strength, in conjunction with Eq. (\ref{eq:zhun}), forms the foundation of our quasi-resonance approach.
	
	Fig.~\ref{fig:zhun}(a) shows $\langle p_4(k)\rangle = \sum_k p_4(k)/N$ as a function of $\Omega$ for various lattice sizes, under two different parameter sets: $b=3$ and $\delta m=0$, and $b=0.1$ and $\delta m=0.05$. Fig.~\ref{fig:can}(a) demonstrates that for the lattice sizes studied here, all modes are extended with $b=3$ and $\delta m=0$, whereas a mix of both extended and localized modes is observed for $b=0.1$ and $\delta m=0.05$. A crucial observation is the uniform reduction of $\langle p_4(k) \rangle$ with decreasing $\Omega$ in a lattice of fixed size. This property suggests that $\gamma_{1}$ tends to zero, resulting in a reduced influence of the 4-wave resonance network when $\Omega$ is sufficiently small, or equivalently, when $g$ is sufficiently low.
	
	In the case of a fixed $\Omega$, we observe a monotonic increase in $\langle p_4(k)\rangle$ with the lattice size $N$. This indicates that for sufficiently large lattices, $\langle p_4(k)\rangle$ might become significant enough to ensure that each mode is adequately connected to a sufficient number of other modes. As a result, we can predict that in the thermodynamic limit, both homogeneous and disordered lattices, if the latter featuring a mix of extended and localized modes, will be predominantly influenced by 4-wave quasi-resonances.
	
	Fig.~\ref{fig:zhun}(b) presents the results for $b=3$ and $\delta m=0.5$. According to Fig.~\ref{fig:can}(a), all modes become localized when the lattice size exceeds $N=100$. Notably, for all three lattice sizes, $\langle p_4(k)\rangle$ collapses onto a single curve, marking a significant departure from the scenarios depicted in Fig.~\ref{fig:zhun}(a). This indicates a size-independent connection strength in completely localized lattices. Considering the monotonic decrease of $\langle p_4(k)\rangle$ with decreasing $\Omega$, it is reasonable to predict that the dominance of 4-wave quasi-resonances diminishes in regions where $g$ is sufficiently small. This behavior can be readily understood: for a localized mode, its $A_{k_1k_2k_3k_4}$ is determined by the mode's localization length, leading to a size-independent $p_4(k)$. Conversely, in an extended mode that spans the entire lattice, its $p_4(k)$ includes an increasing number of resonance sets as $N$ increases.
	
	The distribution of $p_4(k)$ acts as direct evidence for the connectivity within the network. Figs.~\ref{fig:zhun}(c) and \ref{fig:zhun}(d) display the distributions of $p_4(k)$ for two distinct $\Omega$ values, obtained for $N = 16384$ with $b = 3$ and $\delta m = 0.5$. At $\Omega = 0.01$, the distribution exhibits a gap around the origin, $p_4(k)=0$, signifying the absence of isolated modes. The majority of modes have a high $p_4(k)$ value, indicating a significant degree of connectivity within the network. In contrast, at $\Omega = 0.0001$, the distribution shifts towards a power-law, characterized by a pronounced peak at the origin. This shift suggests that a large number of modes do not meet the resonance condition and are therefore isolated from the resonance network. It highlights that in completely localized lattices, a lower $\Omega$ results in disconnection within the 4-wave resonance network.
	
	We can estimate the scaling relationship between the threshold $g_c$, which indicates diminished network connectivity, and $\delta m$. The presence of $p_4(k)$ in the range $[0, 1 \times 10^{-9}]$ is used to signify the emergence of isolated modes and a loss of connectivity in the 4-wave resonance network. It is important to note that varying the width of this interval does not significantly alter the scaling law. Direct measurement reveals that $\Omega_c \sim {\delta m}^{5.1}$. Following this, we analyze the phonon spectrum to determine the frequency broadening $\gamma$ for a specific $g$, leading to the derivation $\gamma \sim g^{1.3}$. Considering that $\Omega$ is proportional to $\gamma$, we subsequently deduce that $g_c \sim {\delta m}^{3.9}$. Detailed explanations are provided in part III of SM~\cite{SM}.
	
	A large $p_4(k)$ indicates the dominance of 4-wave quasi-resonance and ensures a non-vanishing $\gamma_{1}$. In this case, the thermalization time scales as $T_{\text{eq}} \propto g^{-2}$. However, when disconnection occurs, wave turbulence theory suggests a shift towards the predominance of higher-order resonances, thereby adjusting the scaling law to $T_{\text{eq}} \propto g^{-n}$, with $n > 2$~\cite{Lvov:2010,Onorato:2015,Lvov:2018,Pistone:2018,Pistone:2019,Onorato:2023}. Within the context of the $\phi_4$ model, it can be demonstrated (see part II, SM \cite{SM}) that $n$ takes on values such as $4, 6, 8, 10, \ldots$ as $g$ decreases.
	
	\begin{figure}[tp]
		\includegraphics{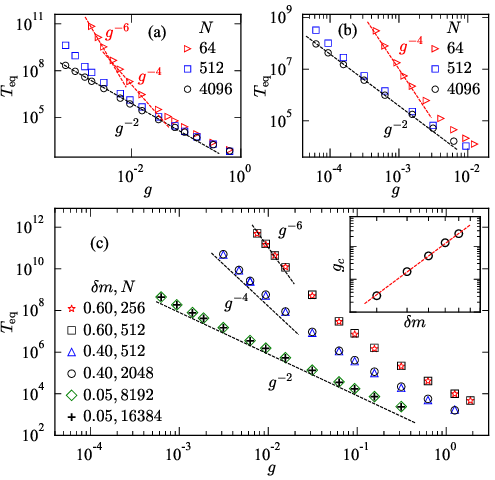}
		\caption{\label{fig:junyun}
			Validation of quasi-resonance predictions. The data is presented for lattices with (a) fully extended modes ($b=3.0$ and $\delta m=0.00$), (b) partially extended modes ($b=0.1$ and $\delta m=0.05$), and (c) completely localized modes ($b=3.0$ and $\delta m=0.05, 0.4, 0.6$) across various lattice sizes. Reference dashed lines are included to illustrate typical slopes. The inset in (c) highlights a dashed line of scaling $g_c \propto \delta m^{3.9}$, which roughly agrees with the measured data (circles).}
	\end{figure}
	
	To validate the quasi-resonance predictions, we numerically computed the thermalization time $T_{\text{eq}}$ using the method outlined in \cite{Onorato:2015, Pistone:2018}. This approach defines entropy as $s(t) = \sum_k f_k \log(f_k)$, with $f_k = E_k/\langle E_k \rangle$ and $\langle E_k \rangle = \sum E_k/N$. As the system approaches equipartition, $E_k$ trends towards $\langle E_k \rangle$, driving $s(t)$ towards zero. In our simulations, each mode receives random initial values, ensuring constant energy density $\epsilon$ through renormalization. By adjusting $\beta$, we vary $g$. For a stabilized entropy function $s(t)$, multiple ensemble runs are executed for each $g$ value. The thermalization time is ascertained using a threshold of $s(t) = 0.1N$. The choice of threshold does not affect the scaling law of $T_{\text{eq}}$~\cite{Pistone:2018,SM}.
	
	As depicted in Fig.~\ref{fig:junyun}, our simulation results fully confirm the predictions. Figs.~\ref{fig:junyun}(a) and (b) illustrate the outcomes for lattices with entirely extended modes and those with a mix of modes, respectively. In both cases, smaller lattices exhibit a transition from 4-wave to higher-order resonance predictions as $g$ decreases, signifying the disconnection of the 4-wave resonance network. In contrast, for larger lattices, the results consistently align with the 4-wave resonance prediction throughout the entire range of $g$ values tested.
	
	Fig.~\ref{fig:junyun}(c) presents the results for lattices with completely localized modes. A striking observation is that, regardless of the lattice size, the thermalization time is nearly identical for a given disorder degree $\delta m$ at the same $g$ value. With $\delta m=0.05$, the scaling predicted by the 4-wave resonance is consistently observed across the entire range of $g$ values we investigated. However, as the disorder degree increases, transitions in thermalization time scaling become apparent. Notably, the scaling shifts from 4-wave to 6-wave resonance predictions at $\delta m = 0.4$ and $\delta m = 0.6$, and further to an 8-wave resonance prediction at $\delta m = 0.6$ with decreasing $g$. The inset of Fig.~\ref{fig:junyun}(c) validates the predicted $g_c \sim {\delta m}^{3.9}$ for the connectivity loss in the 4-wave network, as evidenced by the directly measured data. The correlation is found to be satisfactory.
	
	In conclusion, we have established a general approach to analyze and understand energy diffusion in lattices, emphasizing the role of eigenmode properties over other specific lattice characteristics. The answer to the FPUT problem about whether weak enough perturbations can lead to the equaipartition of lattices in the thermodynamic limit therefore reached: the class of lattices having extended modes can achieve thermalization even with arbitrarily weak perturbations,  consistent with traditional assumptions in classical statistical physics. While the second class with only fully localized modes may not undergo thermalization under similar conditions, thereby establishing theoretical boundaries for the application of classical statistical physics. Our research not only significantly deepens the understanding of the FPUT problem, but also lays a foundation for future research on energy transfer in many-body systems characterized by strong nonlinearity.

	
	\bibliography{ref}

\begin{thebibliography}{44}%
\makeatletter
\providecommand \@ifxundefined [1]{%
 \@ifx{#1\undefined}
}%
\providecommand \@ifnum [1]{%
 \ifnum #1\expandafter \@firstoftwo
 \else \expandafter \@secondoftwo
 \fi
}%
\providecommand \@ifx [1]{%
 \ifx #1\expandafter \@firstoftwo
 \else \expandafter \@secondoftwo
 \fi
}%
\providecommand \natexlab [1]{#1}%
\providecommand \enquote  [1]{``#1''}%
\providecommand \bibnamefont  [1]{#1}%
\providecommand \bibfnamefont [1]{#1}%
\providecommand \citenamefont [1]{#1}%
\providecommand \href@noop [0]{\@secondoftwo}%
\providecommand \href [0]{\begingroup \@sanitize@url \@href}%
\providecommand \@href[1]{\@@startlink{#1}\@@href}%
\providecommand \@@href[1]{\endgroup#1\@@endlink}%
\providecommand \@sanitize@url [0]{\catcode `\\12\catcode `\$12\catcode
  `\&12\catcode `\#12\catcode `\^12\catcode `\_12\catcode `\%12\relax}%
\providecommand \@@startlink[1]{}%
\providecommand \@@endlink[0]{}%
\providecommand \url  [0]{\begingroup\@sanitize@url \@url }%
\providecommand \@url [1]{\endgroup\@href {#1}{\urlprefix }}%
\providecommand \urlprefix  [0]{URL }%
\providecommand \Eprint [0]{\href }%
\providecommand \doibase [0]{https://doi.org/}%
\providecommand \selectlanguage [0]{\@gobble}%
\providecommand \bibinfo  [0]{\@secondoftwo}%
\providecommand \bibfield  [0]{\@secondoftwo}%
\providecommand \translation [1]{[#1]}%
\providecommand \BibitemOpen [0]{}%
\providecommand \bibitemStop [0]{}%
\providecommand \bibitemNoStop [0]{.\EOS\space}%
\providecommand \EOS [0]{\spacefactor3000\relax}%
\providecommand \BibitemShut  [1]{\csname bibitem#1\endcsname}%
\let\auto@bib@innerbib\@empty
\bibitem [{\citenamefont {Fermi}\ \emph {et~al.}()\citenamefont {Fermi},
  \citenamefont {Pasta},\ and\ \citenamefont {Ulam}}]{Fermi:1955}%
  \BibitemOpen
  \bibfield  {author} {\bibinfo {author} {\bibfnamefont {E.}~\bibnamefont
  {Fermi}}, \bibinfo {author} {\bibfnamefont {J.}~\bibnamefont {Pasta}},\ and\
  \bibinfo {author} {\bibfnamefont {S.}~\bibnamefont {Ulam}},\ }\href@noop {}
  {\bibinfo  {journal} {Los Alamos Scientific Laboratory, Report No. LA-1940,
  1955}\ }\BibitemShut {NoStop}%
\bibitem [{\citenamefont {Zabusky}\ and\ \citenamefont
  {Kruskal}(1965)}]{Zabusky:1965}%
  \BibitemOpen
\bibfield  {journal} {  }\bibfield  {author} {\bibinfo {author} {\bibfnamefont
  {N.~J.}\ \bibnamefont {Zabusky}}\ and\ \bibinfo {author} {\bibfnamefont
  {M.~D.}\ \bibnamefont {Kruskal}},\ }\href
  {https://doi.org/10.1103/PhysRevLett.15.240} {\bibfield  {journal} {\bibinfo
  {journal} {Phys. Rev. Lett.}\ }\textbf {\bibinfo {volume} {15}},\ \bibinfo
  {pages} {240} (\bibinfo {year} {1965})}\BibitemShut {NoStop}%
\bibitem [{\citenamefont {Fucito}\ \emph {et~al.}(1982)\citenamefont {Fucito},
  \citenamefont {Marchesoni}, \citenamefont {Marinari}, \citenamefont {Parisi},
  \citenamefont {Peliti}, \citenamefont {Ruffo},\ and\ \citenamefont
  {Vulpiani}}]{Fucito:1982}%
  \BibitemOpen
  \bibfield  {author} {\bibinfo {author} {\bibfnamefont {F.}~\bibnamefont
  {Fucito}}, \bibinfo {author} {\bibfnamefont {F.}~\bibnamefont {Marchesoni}},
  \bibinfo {author} {\bibfnamefont {E.}~\bibnamefont {Marinari}}, \bibinfo
  {author} {\bibfnamefont {G.}~\bibnamefont {Parisi}}, \bibinfo {author}
  {\bibfnamefont {L.}~\bibnamefont {Peliti}}, \bibinfo {author} {\bibfnamefont
  {S.}~\bibnamefont {Ruffo}},\ and\ \bibinfo {author} {\bibfnamefont
  {A.}~\bibnamefont {Vulpiani}},\ }\href
  {https://doi.org/https://doi.org/10.1051/jphys:01982004305070700} {\bibfield
  {journal} {\bibinfo  {journal} {J. Phys. (Les Ulis, Fr.)}\ }\textbf {\bibinfo
  {volume} {43}},\ \bibinfo {pages} {707} (\bibinfo {year} {1982})}\BibitemShut
  {NoStop}%
\bibitem [{\citenamefont {De~Luca}\ \emph {et~al.}(1995)\citenamefont
  {De~Luca}, \citenamefont {Lichtenberg},\ and\ \citenamefont
  {Ruffo}}]{Luca:1995}%
  \BibitemOpen
  \bibfield  {author} {\bibinfo {author} {\bibfnamefont {J.}~\bibnamefont
  {De~Luca}}, \bibinfo {author} {\bibfnamefont {A.~J.}\ \bibnamefont
  {Lichtenberg}},\ and\ \bibinfo {author} {\bibfnamefont {S.}~\bibnamefont
  {Ruffo}},\ }\href {https://doi.org/10.1103/PhysRevE.51.2877} {\bibfield
  {journal} {\bibinfo  {journal} {Phys. Rev. E}\ }\textbf {\bibinfo {volume}
  {51}},\ \bibinfo {pages} {2877} (\bibinfo {year} {1995})}\BibitemShut
  {NoStop}%
\bibitem [{\citenamefont {Poggi}\ \emph {et~al.}(1995)\citenamefont {Poggi},
  \citenamefont {Ruffo},\ and\ \citenamefont {Kantz}}]{Poggi:1995}%
  \BibitemOpen
  \bibfield  {author} {\bibinfo {author} {\bibfnamefont {P.}~\bibnamefont
  {Poggi}}, \bibinfo {author} {\bibfnamefont {S.}~\bibnamefont {Ruffo}},\ and\
  \bibinfo {author} {\bibfnamefont {H.}~\bibnamefont {Kantz}},\ }\href
  {https://doi.org/10.1103/PhysRevE.52.307} {\bibfield  {journal} {\bibinfo
  {journal} {Phys. Rev. E}\ }\textbf {\bibinfo {volume} {52}},\ \bibinfo
  {pages} {307} (\bibinfo {year} {1995})}\BibitemShut {NoStop}%
\bibitem [{\citenamefont {Parisi}(1997)}]{Parisi:1997}%
  \BibitemOpen
  \bibfield  {author} {\bibinfo {author} {\bibfnamefont {G.}~\bibnamefont
  {Parisi}},\ }\href {https://doi.org/10.1209/epl/i1997-00471-9} {\bibfield
  {journal} {\bibinfo  {journal} {Europhys. Lett.}\ }\textbf {\bibinfo {volume}
  {40}},\ \bibinfo {pages} {357} (\bibinfo {year} {1997})}\BibitemShut
  {NoStop}%
\bibitem [{\citenamefont {Casetti}\ \emph {et~al.}(1997)\citenamefont
  {Casetti}, \citenamefont {Cerruti-Sola}, \citenamefont {Pettini},\ and\
  \citenamefont {Cohen}}]{Casetti:1997}%
  \BibitemOpen
  \bibfield  {author} {\bibinfo {author} {\bibfnamefont {L.}~\bibnamefont
  {Casetti}}, \bibinfo {author} {\bibfnamefont {M.}~\bibnamefont
  {Cerruti-Sola}}, \bibinfo {author} {\bibfnamefont {M.}~\bibnamefont
  {Pettini}},\ and\ \bibinfo {author} {\bibfnamefont {E.~G.~D.}\ \bibnamefont
  {Cohen}},\ }\href {https://doi.org/10.1103/PhysRevE.55.6566} {\bibfield
  {journal} {\bibinfo  {journal} {Phys. Rev. E}\ }\textbf {\bibinfo {volume}
  {55}},\ \bibinfo {pages} {6566} (\bibinfo {year} {1997})}\BibitemShut
  {NoStop}%
\bibitem [{\citenamefont {De~Luca}\ \emph {et~al.}(1999)\citenamefont
  {De~Luca}, \citenamefont {Lichtenberg},\ and\ \citenamefont
  {Ruffo}}]{Luca:1999}%
  \BibitemOpen
  \bibfield  {author} {\bibinfo {author} {\bibfnamefont {J.}~\bibnamefont
  {De~Luca}}, \bibinfo {author} {\bibfnamefont {A.~J.}\ \bibnamefont
  {Lichtenberg}},\ and\ \bibinfo {author} {\bibfnamefont {S.}~\bibnamefont
  {Ruffo}},\ }\href {https://doi.org/10.1103/PhysRevE.60.3781} {\bibfield
  {journal} {\bibinfo  {journal} {Phys. Rev. E}\ }\textbf {\bibinfo {volume}
  {60}},\ \bibinfo {pages} {3781} (\bibinfo {year} {1999})}\BibitemShut
  {NoStop}%
\bibitem [{\citenamefont {Berchialla}\ \emph {et~al.}(2004)\citenamefont
  {Berchialla}, \citenamefont {Giorgilli},\ and\ \citenamefont
  {Paleari}}]{Berchialla:2004}%
  \BibitemOpen
  \bibfield  {author} {\bibinfo {author} {\bibfnamefont {L.}~\bibnamefont
  {Berchialla}}, \bibinfo {author} {\bibfnamefont {A.}~\bibnamefont
  {Giorgilli}},\ and\ \bibinfo {author} {\bibfnamefont {S.}~\bibnamefont
  {Paleari}},\ }\href
  {https://doi.org/https://doi.org/10.1016/j.physleta.2003.11.052} {\bibfield
  {journal} {\bibinfo  {journal} {Phys. Lett. A}\ }\textbf {\bibinfo {volume}
  {321}},\ \bibinfo {pages} {167} (\bibinfo {year} {2004})}\BibitemShut
  {NoStop}%
\bibitem [{\citenamefont {Gershgorin}\ \emph {et~al.}(2007)\citenamefont
  {Gershgorin}, \citenamefont {Lvov},\ and\ \citenamefont
  {Cai}}]{Gershgorin:2007}%
  \BibitemOpen
  \bibfield  {author} {\bibinfo {author} {\bibfnamefont {B.}~\bibnamefont
  {Gershgorin}}, \bibinfo {author} {\bibfnamefont {Y.~V.}\ \bibnamefont
  {Lvov}},\ and\ \bibinfo {author} {\bibfnamefont {D.}~\bibnamefont {Cai}},\
  }\href {https://doi.org/10.1103/PhysRevE.75.046603} {\bibfield  {journal}
  {\bibinfo  {journal} {Phys. Rev. E}\ }\textbf {\bibinfo {volume} {75}},\
  \bibinfo {pages} {046603} (\bibinfo {year} {2007})}\BibitemShut {NoStop}%
\bibitem [{\citenamefont {Gallavotti}(2007)}]{Gallavotti:2008}%
  \BibitemOpen
  \bibfield  {author} {\bibinfo {author} {\bibfnamefont {G.}~\bibnamefont
  {Gallavotti}},\ }\href@noop {} {\emph {\bibinfo {title} {The Fermi-Pasta-Ulam
  problem: a status report}}},\ Vol.\ \bibinfo {volume} {728}\ (\bibinfo
  {publisher} {Springer},\ \bibinfo {address} {New York},\ \bibinfo {year}
  {2007})\BibitemShut {NoStop}%
\bibitem [{\citenamefont {Benettin}\ \emph {et~al.}(2009)\citenamefont
  {Benettin}, \citenamefont {Livi},\ and\ \citenamefont
  {Ponno}}]{Benettin:2009}%
  \BibitemOpen
  \bibfield  {author} {\bibinfo {author} {\bibfnamefont {G.}~\bibnamefont
  {Benettin}}, \bibinfo {author} {\bibfnamefont {R.}~\bibnamefont {Livi}},\
  and\ \bibinfo {author} {\bibfnamefont {A.}~\bibnamefont {Ponno}},\ }\href
  {https://doi.org/10.1007/s10955-008-9660-6} {\bibfield  {journal} {\bibinfo
  {journal} {J. Stat. Phys.}\ }\textbf {\bibinfo {volume} {135}},\ \bibinfo
  {pages} {873} (\bibinfo {year} {2009})}\BibitemShut {NoStop}%
\bibitem [{\citenamefont {Cretegny}\ \emph {et~al.}(1998)\citenamefont
  {Cretegny}, \citenamefont {Dauxois}, \citenamefont {Ruffo},\ and\
  \citenamefont {Torcini}}]{Cretegny:1998}%
  \BibitemOpen
  \bibfield  {author} {\bibinfo {author} {\bibfnamefont {T.}~\bibnamefont
  {Cretegny}}, \bibinfo {author} {\bibfnamefont {T.}~\bibnamefont {Dauxois}},
  \bibinfo {author} {\bibfnamefont {S.}~\bibnamefont {Ruffo}},\ and\ \bibinfo
  {author} {\bibfnamefont {A.}~\bibnamefont {Torcini}},\ }\href
  {https://doi.org/10.1016/S0167-2789(98)00107-9} {\bibfield  {journal}
  {\bibinfo  {journal} {Physica D}\ }\textbf {\bibinfo {volume} {121}},\
  \bibinfo {pages} {109} (\bibinfo {year} {1998})}\BibitemShut {NoStop}%
\bibitem [{\citenamefont {Lepri}(1998)}]{Lepri:1998}%
  \BibitemOpen
  \bibfield  {author} {\bibinfo {author} {\bibfnamefont {S.}~\bibnamefont
  {Lepri}},\ }\href {https://doi.org/10.1103/PhysRevE.58.7165} {\bibfield
  {journal} {\bibinfo  {journal} {Phys. Rev. E}\ }\textbf {\bibinfo {volume}
  {58}},\ \bibinfo {pages} {7165} (\bibinfo {year} {1998})}\BibitemShut
  {NoStop}%
\bibitem [{\citenamefont {Benettin}\ and\ \citenamefont
  {Ponno}(2011)}]{Benettin:2011}%
  \BibitemOpen
  \bibfield  {author} {\bibinfo {author} {\bibfnamefont {G.}~\bibnamefont
  {Benettin}}\ and\ \bibinfo {author} {\bibfnamefont {A.}~\bibnamefont
  {Ponno}},\ }\href {https://doi.org/10.1007/s10955-011-0277-9} {\bibfield
  {journal} {\bibinfo  {journal} {J. Stat. Phys.}\ }\textbf {\bibinfo {volume}
  {144}},\ \bibinfo {pages} {793} (\bibinfo {year} {2011})}\BibitemShut
  {NoStop}%
\bibitem [{\citenamefont {Benettin}\ \emph {et~al.}(2013)\citenamefont
  {Benettin}, \citenamefont {Christodoulidi},\ and\ \citenamefont
  {Ponno}}]{Benettin:2013}%
  \BibitemOpen
  \bibfield  {author} {\bibinfo {author} {\bibfnamefont {G.}~\bibnamefont
  {Benettin}}, \bibinfo {author} {\bibfnamefont {H.}~\bibnamefont
  {Christodoulidi}},\ and\ \bibinfo {author} {\bibfnamefont {A.}~\bibnamefont
  {Ponno}},\ }\href {https://doi.org/https://doi.org/10.1007/s10955-013-0760-6}
  {\bibfield  {journal} {\bibinfo  {journal} {J. Stat. Phys.}\ }\textbf
  {\bibinfo {volume} {152}},\ \bibinfo {pages} {195} (\bibinfo {year}
  {2013})}\BibitemShut {NoStop}%
\bibitem [{\citenamefont {Zhang}\ \emph {et~al.}(2016)\citenamefont {Zhang},
  \citenamefont {Tang},\ and\ \citenamefont {Tong}}]{Zhang:2016}%
  \BibitemOpen
  \bibfield  {author} {\bibinfo {author} {\bibfnamefont {Z.}~\bibnamefont
  {Zhang}}, \bibinfo {author} {\bibfnamefont {C.}~\bibnamefont {Tang}},\ and\
  \bibinfo {author} {\bibfnamefont {P.}~\bibnamefont {Tong}},\ }\href
  {https://doi.org/10.1103/PhysRevE.93.022216} {\bibfield  {journal} {\bibinfo
  {journal} {Phys. Rev. E}\ }\textbf {\bibinfo {volume} {93}},\ \bibinfo
  {pages} {022216} (\bibinfo {year} {2016})}\BibitemShut {NoStop}%
\bibitem [{\citenamefont {Sun}\ \emph {et~al.}(2020)\citenamefont {Sun},
  \citenamefont {Zhang},\ and\ \citenamefont {Tong}}]{Sun:2020}%
  \BibitemOpen
  \bibfield  {author} {\bibinfo {author} {\bibfnamefont {L.}~\bibnamefont
  {Sun}}, \bibinfo {author} {\bibfnamefont {Z.}~\bibnamefont {Zhang}},\ and\
  \bibinfo {author} {\bibfnamefont {P.}~\bibnamefont {Tong}},\ }\href
  {https://doi.org/10.1088/1367-2630/ab9770} {\bibfield  {journal} {\bibinfo
  {journal} {New J. Phys.}\ }\textbf {\bibinfo {volume} {22}},\ \bibinfo
  {pages} {073027} (\bibinfo {year} {2020})}\BibitemShut {NoStop}%
\bibitem [{\citenamefont {Ganapa}\ \emph {et~al.}(2020)\citenamefont {Ganapa},
  \citenamefont {Apte},\ and\ \citenamefont {Dhar}}]{Ganapa:2020}%
  \BibitemOpen
  \bibfield  {author} {\bibinfo {author} {\bibfnamefont {S.}~\bibnamefont
  {Ganapa}}, \bibinfo {author} {\bibfnamefont {A.}~\bibnamefont {Apte}},\ and\
  \bibinfo {author} {\bibfnamefont {A.}~\bibnamefont {Dhar}},\ }\href
  {https://doi.org/https://doi.org/10.1007/s10955-020-02576-2} {\bibfield
  {journal} {\bibinfo  {journal} {J. Stat. Phys.}\ }\textbf {\bibinfo {volume}
  {180}},\ \bibinfo {pages} {1010} (\bibinfo {year} {2020})}\BibitemShut
  {NoStop}%
\bibitem [{\citenamefont {Onorato}\ \emph {et~al.}(2015)\citenamefont
  {Onorato}, \citenamefont {Vozella}, \citenamefont {Proment},\ and\
  \citenamefont {Lvov}}]{Onorato:2015}%
  \BibitemOpen
  \bibfield  {author} {\bibinfo {author} {\bibfnamefont {M.}~\bibnamefont
  {Onorato}}, \bibinfo {author} {\bibfnamefont {L.}~\bibnamefont {Vozella}},
  \bibinfo {author} {\bibfnamefont {D.}~\bibnamefont {Proment}},\ and\ \bibinfo
  {author} {\bibfnamefont {Y.~V.}\ \bibnamefont {Lvov}},\ }\href
  {https://doi.org/10.1073/pnas.1404397112} {\bibfield  {journal} {\bibinfo
  {journal} {Proc. Natl. Acad. Sci. U.S.A.}\ }\textbf {\bibinfo {volume}
  {112}},\ \bibinfo {pages} {4208} (\bibinfo {year} {2015})}\BibitemShut
  {NoStop}%
\bibitem [{\citenamefont {Lvov}\ and\ \citenamefont
  {Onorato}(2018)}]{Lvov:2018}%
  \BibitemOpen
  \bibfield  {author} {\bibinfo {author} {\bibfnamefont {Y.~V.}\ \bibnamefont
  {Lvov}}\ and\ \bibinfo {author} {\bibfnamefont {M.}~\bibnamefont {Onorato}},\
  }\href {https://doi.org/10.1103/PhysRevLett.120.144301} {\bibfield  {journal}
  {\bibinfo  {journal} {Phys. Rev. Lett.}\ }\textbf {\bibinfo {volume} {120}},\
  \bibinfo {pages} {144301} (\bibinfo {year} {2018})}\BibitemShut {NoStop}%
\bibitem [{\citenamefont {Pistone}\ \emph {et~al.}(2018)\citenamefont
  {Pistone}, \citenamefont {Onorato},\ and\ \citenamefont
  {Chibbaro}}]{Pistone:2018}%
  \BibitemOpen
  \bibfield  {author} {\bibinfo {author} {\bibfnamefont {L.}~\bibnamefont
  {Pistone}}, \bibinfo {author} {\bibfnamefont {M.}~\bibnamefont {Onorato}},\
  and\ \bibinfo {author} {\bibfnamefont {S.}~\bibnamefont {Chibbaro}},\ }\href
  {https://doi.org/10.1209/0295-5075/121/44003} {\bibfield  {journal} {\bibinfo
   {journal} {Europhys. Lett.}\ }\textbf {\bibinfo {volume} {121}},\ \bibinfo
  {pages} {44003} (\bibinfo {year} {2018})}\BibitemShut {NoStop}%
\bibitem [{\citenamefont {Fu}\ \emph {et~al.}(2019{\natexlab{a}})\citenamefont
  {Fu}, \citenamefont {Zhang},\ and\ \citenamefont {Zhao}}]{Fu:2019R}%
  \BibitemOpen
  \bibfield  {author} {\bibinfo {author} {\bibfnamefont {W.}~\bibnamefont
  {Fu}}, \bibinfo {author} {\bibfnamefont {Y.}~\bibnamefont {Zhang}},\ and\
  \bibinfo {author} {\bibfnamefont {H.}~\bibnamefont {Zhao}},\ }\href
  {https://doi.org/10.1103/PhysRevE.100.010101} {\bibfield  {journal} {\bibinfo
   {journal} {Phys. Rev. E}\ }\textbf {\bibinfo {volume} {100}},\ \bibinfo
  {pages} {010101(R)} (\bibinfo {year} {2019}{\natexlab{a}})}\BibitemShut
  {NoStop}%
\bibitem [{\citenamefont {Fu}\ \emph {et~al.}(2019{\natexlab{b}})\citenamefont
  {Fu}, \citenamefont {Zhang},\ and\ \citenamefont {Zhao}}]{Fu:2019}%
  \BibitemOpen
  \bibfield  {author} {\bibinfo {author} {\bibfnamefont {W.}~\bibnamefont
  {Fu}}, \bibinfo {author} {\bibfnamefont {Y.}~\bibnamefont {Zhang}},\ and\
  \bibinfo {author} {\bibfnamefont {H.}~\bibnamefont {Zhao}},\ }\href
  {https://doi.org/10.1088/1367-2630/ab115a} {\bibfield  {journal} {\bibinfo
  {journal} {New J. Phys.}\ }\textbf {\bibinfo {volume} {21}},\ \bibinfo
  {pages} {043009} (\bibinfo {year} {2019}{\natexlab{b}})}\BibitemShut
  {NoStop}%
\bibitem [{\citenamefont {Pistone}\ \emph {et~al.}(2019)\citenamefont
  {Pistone}, \citenamefont {Chibbaro}, \citenamefont {Bustamante},
  \citenamefont {Lvov},\ and\ \citenamefont {Onorato}}]{Pistone:2019}%
  \BibitemOpen
  \bibfield  {author} {\bibinfo {author} {\bibfnamefont {L.}~\bibnamefont
  {Pistone}}, \bibinfo {author} {\bibfnamefont {S.}~\bibnamefont {Chibbaro}},
  \bibinfo {author} {\bibfnamefont {M.~D.}\ \bibnamefont {Bustamante}},
  \bibinfo {author} {\bibfnamefont {Y.}~\bibnamefont {Lvov}, \bibfnamefont
  {V}},\ and\ \bibinfo {author} {\bibfnamefont {M.}~\bibnamefont {Onorato}},\
  }\href {https://doi.org/10.3934/mine.2019.4.672} {\bibfield  {journal}
  {\bibinfo  {journal} {Math. Eng.}\ }\textbf {\bibinfo {volume} {1}},\
  \bibinfo {pages} {672} (\bibinfo {year} {2019})}\BibitemShut {NoStop}%
\bibitem [{\citenamefont {Onorato}\ \emph {et~al.}(2023)\citenamefont
  {Onorato}, \citenamefont {Lvov}, \citenamefont {Dematteis},\ and\
  \citenamefont {Chibbaro}}]{Onorato:2023}%
  \BibitemOpen
  \bibfield  {author} {\bibinfo {author} {\bibfnamefont {M.}~\bibnamefont
  {Onorato}}, \bibinfo {author} {\bibfnamefont {Y.}~\bibnamefont {Lvov}},
  \bibinfo {author} {\bibfnamefont {G.}~\bibnamefont {Dematteis}},\ and\
  \bibinfo {author} {\bibfnamefont {S.}~\bibnamefont {Chibbaro}},\ }\href
  {https://doi.org/https://doi.org/10.1016/j.physrep.2023.09.006} {\bibfield
  {journal} {\bibinfo  {journal} {Phys. Rep.}\ }\textbf {\bibinfo {volume}
  {1040}},\ \bibinfo {pages} {1} (\bibinfo {year} {2023})}\BibitemShut
  {NoStop}%
\bibitem [{\citenamefont {Wang}\ \emph {et~al.}(2020)\citenamefont {Wang},
  \citenamefont {Fu}, \citenamefont {Zhang},\ and\ \citenamefont
  {Zhao}}]{Wang:2020}%
  \BibitemOpen
  \bibfield  {author} {\bibinfo {author} {\bibfnamefont {Z.}~\bibnamefont
  {Wang}}, \bibinfo {author} {\bibfnamefont {W.}~\bibnamefont {Fu}}, \bibinfo
  {author} {\bibfnamefont {Y.}~\bibnamefont {Zhang}},\ and\ \bibinfo {author}
  {\bibfnamefont {H.}~\bibnamefont {Zhao}},\ }\href
  {https://doi.org/10.1103/PhysRevLett.124.186401} {\bibfield  {journal}
  {\bibinfo  {journal} {Phys. Rev. Lett.}\ }\textbf {\bibinfo {volume} {124}},\
  \bibinfo {pages} {186401} (\bibinfo {year} {2020})}\BibitemShut {NoStop}%
\bibitem [{\citenamefont {Wang}\ \emph {et~al.}()\citenamefont {Wang},
  \citenamefont {Fu}, \citenamefont {Zhang},\ and\ \citenamefont
  {Zhao}}]{Wang:2020a}%
  \BibitemOpen
  \bibfield  {author} {\bibinfo {author} {\bibfnamefont {Z.}~\bibnamefont
  {Wang}}, \bibinfo {author} {\bibfnamefont {W.}~\bibnamefont {Fu}}, \bibinfo
  {author} {\bibfnamefont {Y.}~\bibnamefont {Zhang}},\ and\ \bibinfo {author}
  {\bibfnamefont {H.}~\bibnamefont {Zhao}},\ }\href@noop {} {}\Eprint
  {https://arxiv.org/abs/2005.03478} {arXiv:2005.03478} \BibitemShut {NoStop}%
\bibitem [{\citenamefont {Zakharov}\ \emph {et~al.}(2004)\citenamefont
  {Zakharov}, \citenamefont {Dias},\ and\ \citenamefont
  {Pushkarev}}]{Zakharov:2004}%
  \BibitemOpen
  \bibfield  {author} {\bibinfo {author} {\bibfnamefont {V.}~\bibnamefont
  {Zakharov}}, \bibinfo {author} {\bibfnamefont {F.}~\bibnamefont {Dias}},\
  and\ \bibinfo {author} {\bibfnamefont {A.}~\bibnamefont {Pushkarev}},\ }\href
  {https://doi.org/https://doi.org/10.1016/j.physrep.2004.04.002} {\bibfield
  {journal} {\bibinfo  {journal} {Phys. Rep.}\ }\textbf {\bibinfo {volume}
  {398}},\ \bibinfo {pages} {1} (\bibinfo {year} {2004})}\BibitemShut {NoStop}%
\bibitem [{\citenamefont {Nazarenko}(2011)}]{Nazarenko:2011}%
  \BibitemOpen
  \bibfield  {author} {\bibinfo {author} {\bibfnamefont {S.}~\bibnamefont
  {Nazarenko}},\ }\href@noop {} {\emph {\bibinfo {title} {Wave turbulence}}},\
  Vol.\ \bibinfo {volume} {825}\ (\bibinfo  {publisher} {Springer,Berlin},\
  \bibinfo {year} {2011})\BibitemShut {NoStop}%
\bibitem [{\citenamefont {Connaughton}\ \emph {et~al.}(2001)\citenamefont
  {Connaughton}, \citenamefont {Nazarenko},\ and\ \citenamefont
  {Pushkarev}}]{Connaughton:2001}%
  \BibitemOpen
  \bibfield  {author} {\bibinfo {author} {\bibfnamefont {C.}~\bibnamefont
  {Connaughton}}, \bibinfo {author} {\bibfnamefont {S.}~\bibnamefont
  {Nazarenko}},\ and\ \bibinfo {author} {\bibfnamefont {A.}~\bibnamefont
  {Pushkarev}},\ }\href {https://doi.org/10.1103/PhysRevE.63.046306} {\bibfield
   {journal} {\bibinfo  {journal} {Phys. Rev. E}\ }\textbf {\bibinfo {volume}
  {63}},\ \bibinfo {pages} {046306} (\bibinfo {year} {2001})}\BibitemShut
  {NoStop}%
\bibitem [{\citenamefont {Kartashova}(2007)}]{Kartashova:2007}%
  \BibitemOpen
  \bibfield  {author} {\bibinfo {author} {\bibfnamefont {E.}~\bibnamefont
  {Kartashova}},\ }\href {https://doi.org/10.1103/PhysRevLett.98.214502}
  {\bibfield  {journal} {\bibinfo  {journal} {Phys. Rev. Lett.}\ }\textbf
  {\bibinfo {volume} {98}},\ \bibinfo {pages} {214502} (\bibinfo {year}
  {2007})}\BibitemShut {NoStop}%
\bibitem [{\citenamefont {L'vov}\ and\ \citenamefont
  {Nazarenko}(2010)}]{Lvov:2010}%
  \BibitemOpen
  \bibfield  {author} {\bibinfo {author} {\bibfnamefont {V.~S.}\ \bibnamefont
  {L'vov}}\ and\ \bibinfo {author} {\bibfnamefont {S.}~\bibnamefont
  {Nazarenko}},\ }\href {https://doi.org/10.1103/PhysRevE.82.056322} {\bibfield
   {journal} {\bibinfo  {journal} {Phys. Rev. E}\ }\textbf {\bibinfo {volume}
  {82}},\ \bibinfo {pages} {056322} (\bibinfo {year} {2010})}\BibitemShut
  {NoStop}%
\bibitem [{\citenamefont {Alfimov}\ \emph {et~al.}(2014)\citenamefont
  {Alfimov}, \citenamefont {Medvedeva},\ and\ \citenamefont
  {Pelinovsky}}]{Alfimov:2014}%
  \BibitemOpen
  \bibfield  {author} {\bibinfo {author} {\bibfnamefont {G.~L.}\ \bibnamefont
  {Alfimov}}, \bibinfo {author} {\bibfnamefont {E.~V.}\ \bibnamefont
  {Medvedeva}},\ and\ \bibinfo {author} {\bibfnamefont {D.~E.}\ \bibnamefont
  {Pelinovsky}},\ }\href {https://doi.org/10.1103/PhysRevLett.112.054103}
  {\bibfield  {journal} {\bibinfo  {journal} {Phys. Rev. Lett.}\ }\textbf
  {\bibinfo {volume} {112}},\ \bibinfo {pages} {054103} (\bibinfo {year}
  {2014})}\BibitemShut {NoStop}%
\bibitem [{\citenamefont {Kevrekidis}\ \emph {et~al.}(2016)\citenamefont
  {Kevrekidis}, \citenamefont {Cuevas-Maraver},\ and\ \citenamefont
  {Pelinovsky}}]{Kevrekidis:2016}%
  \BibitemOpen
  \bibfield  {author} {\bibinfo {author} {\bibfnamefont {P.~G.}\ \bibnamefont
  {Kevrekidis}}, \bibinfo {author} {\bibfnamefont {J.}~\bibnamefont
  {Cuevas-Maraver}},\ and\ \bibinfo {author} {\bibfnamefont {D.~E.}\
  \bibnamefont {Pelinovsky}},\ }\href
  {https://doi.org/10.1103/PhysRevLett.117.094101} {\bibfield  {journal}
  {\bibinfo  {journal} {Phys. Rev. Lett.}\ }\textbf {\bibinfo {volume} {117}},\
  \bibinfo {pages} {094101} (\bibinfo {year} {2016})}\BibitemShut {NoStop}%
\bibitem [{\citenamefont {Xiong}\ and\ \citenamefont
  {Zhang}(2018)}]{Xiong:2018}%
  \BibitemOpen
  \bibfield  {author} {\bibinfo {author} {\bibfnamefont {D.}~\bibnamefont
  {Xiong}}\ and\ \bibinfo {author} {\bibfnamefont {Y.}~\bibnamefont {Zhang}},\
  }\href {https://doi.org/10.1103/PhysRevE.98.012130} {\bibfield  {journal}
  {\bibinfo  {journal} {Phys. Rev. E}\ }\textbf {\bibinfo {volume} {98}},\
  \bibinfo {pages} {012130} (\bibinfo {year} {2018})}\BibitemShut {NoStop}%
\bibitem [{\citenamefont {Manton}\ \emph {et~al.}(2021)\citenamefont {Manton},
  \citenamefont {Ole\ifmmode~\acute{s}\else \'{s}\fi{}}, \citenamefont
  {Roma\ifmmode~\acute{n}\else \'{n}\fi{}czukiewicz},\ and\ \citenamefont
  {Wereszczy\ifmmode~\acute{n}\else \'{n}\fi{}ski}}]{Manton:2021}%
  \BibitemOpen
  \bibfield  {author} {\bibinfo {author} {\bibfnamefont {N.~S.}\ \bibnamefont
  {Manton}}, \bibinfo {author} {\bibfnamefont {K.}~\bibnamefont
  {Ole\ifmmode~\acute{s}\else \'{s}\fi{}}}, \bibinfo {author} {\bibfnamefont
  {T.}~\bibnamefont {Roma\ifmmode~\acute{n}\else \'{n}\fi{}czukiewicz}},\ and\
  \bibinfo {author} {\bibfnamefont {A.}~\bibnamefont
  {Wereszczy\ifmmode~\acute{n}\else \'{n}\fi{}ski}},\ }\href
  {https://doi.org/10.1103/PhysRevLett.127.071601} {\bibfield  {journal}
  {\bibinfo  {journal} {Phys. Rev. Lett.}\ }\textbf {\bibinfo {volume} {127}},\
  \bibinfo {pages} {071601} (\bibinfo {year} {2021})}\BibitemShut {NoStop}%
\bibitem [{\citenamefont {Goldhirsch}\ \emph {et~al.}(1991)\citenamefont
  {Goldhirsch}, \citenamefont {Lubin},\ and\ \citenamefont
  {Gefen}}]{Goldhirsch:1991}%
  \BibitemOpen
  \bibfield  {author} {\bibinfo {author} {\bibfnamefont {I.}~\bibnamefont
  {Goldhirsch}}, \bibinfo {author} {\bibfnamefont {D.}~\bibnamefont {Lubin}},\
  and\ \bibinfo {author} {\bibfnamefont {Y.}~\bibnamefont {Gefen}},\ }\href
  {https://doi.org/10.1103/PhysRevLett.67.3582} {\bibfield  {journal} {\bibinfo
   {journal} {Phys. Rev. Lett.}\ }\textbf {\bibinfo {volume} {67}},\ \bibinfo
  {pages} {3582} (\bibinfo {year} {1991})}\BibitemShut {NoStop}%
\bibitem [{\citenamefont {Leitner}(2001)}]{Leitner:2001}%
  \BibitemOpen
  \bibfield  {author} {\bibinfo {author} {\bibfnamefont {D.~M.}\ \bibnamefont
  {Leitner}},\ }\href {https://doi.org/10.1103/PhysRevB.64.094201} {\bibfield
  {journal} {\bibinfo  {journal} {Phys. Rev. B}\ }\textbf {\bibinfo {volume}
  {64}},\ \bibinfo {pages} {094201} (\bibinfo {year} {2001})}\BibitemShut
  {NoStop}%
\bibitem [{\citenamefont {Girvin}\ and\ \citenamefont
  {Yang}(2019)}]{Girvin:2019}%
  \BibitemOpen
  \bibfield  {author} {\bibinfo {author} {\bibfnamefont {S.}~\bibnamefont
  {Girvin}}\ and\ \bibinfo {author} {\bibfnamefont {K.}~\bibnamefont {Yang}},\
  }\href@noop {} {\emph {\bibinfo {title} {Modern Condensed Matter Physics}}}\
  (\bibinfo  {publisher} {Cambridge University Press, Cambridge, England},\
  \bibinfo {year} {2019})\BibitemShut {NoStop}%
\bibitem [{SM()}]{SM}%
  \BibitemOpen
  \href@noop {} {\bibinfo {title} {See supplemental material at xx for detailed
  derivations and descriptions}}\BibitemShut {NoStop}%
\bibitem [{\citenamefont {Kramer}\ and\ \citenamefont
  {MacKinnon}(1993)}]{Kramer:1993}%
  \BibitemOpen
  \bibfield  {author} {\bibinfo {author} {\bibfnamefont {B.}~\bibnamefont
  {Kramer}}\ and\ \bibinfo {author} {\bibfnamefont {A.}~\bibnamefont
  {MacKinnon}},\ }\href {https://doi.org/10.1088/0034-4885/56/12/001}
  {\bibfield  {journal} {\bibinfo  {journal} {Rep. Prog. Phys}\ }\textbf
  {\bibinfo {volume} {56}},\ \bibinfo {pages} {1469} (\bibinfo {year}
  {1993})}\BibitemShut {NoStop}%
\bibitem [{\citenamefont {Ishii}(1973)}]{Ishii:1973}%
  \BibitemOpen
  \bibfield  {author} {\bibinfo {author} {\bibfnamefont {K.}~\bibnamefont
  {Ishii}},\ }\href {https://doi.org/10.1143/PTPS.53.77} {\bibfield  {journal}
  {\bibinfo  {journal} {Prog. Theor. Phys. Suppl.}\ }\textbf {\bibinfo {volume}
  {53}},\ \bibinfo {pages} {77} (\bibinfo {year} {1973})}\BibitemShut {NoStop}%
\bibitem [{\citenamefont {Krimer}\ and\ \citenamefont
  {Flach}(2010)}]{Krimer:2010}%
  \BibitemOpen
  \bibfield  {author} {\bibinfo {author} {\bibfnamefont {D.~O.}\ \bibnamefont
  {Krimer}}\ and\ \bibinfo {author} {\bibfnamefont {S.}~\bibnamefont {Flach}},\
  }\href {https://doi.org/10.1103/PhysRevE.82.046221} {\bibfield  {journal}
  {\bibinfo  {journal} {Phys. Rev. E}\ }\textbf {\bibinfo {volume} {82}},\
  \bibinfo {pages} {046221} (\bibinfo {year} {2010})}\BibitemShut {NoStop}%
\end{thebibliography}%
\end{document}